\documentclass[manuscript]{aastex63}
\usepackage{natbib}
\usepackage{fontenc}
\usepackage{comment}
\usepackage{amsmath}
\usepackage{enumitem}
\usepackage{float}
\graphicspath{{./}{Figures/}}

\newcommand{\re}{\textcolor[rgb]{0,0,0}}
\newcommand{\rd}{\textcolor[rgb]{0,0,0}}
\newcommand{\bl}{\textcolor[rgb]{0,0,0}}

\newcommand{\bx}{\textcolor[rgb]{0,0,0}}

\def\gtsima{$\;\buildrel > \over \sim \;$}
\def\simgt{\lower.5ex \hbox{\gtsima}}
\def\ltsima{$\;\buildrel < \over \sim \;$}
\def\simlt{\lower.5ex \hbox{\ltsima}}
\def \CHp{\ifmmode{\rm CH^+}\else{$\rm CH^+$}\fi}
\def \HH{\ifmmode{\rm H_2}\else{$\rm H_2$}\fi}
\def \Cp{\ifmmode{\rm C^+}\else{$\rm C^+$}\fi} 
\def \cc    {\ifmmode{\,{\rm cm}^{-3}}\else{$\,{\rm cm}^{-3}$}\fi}
\def \dens{\ifmmode{n_{\rm H}}\else{$n_{\rm H}$}\fi}
\def \kms   {\ifmmode{\,{\rm km}\,{\rm s}^{-1}}\else{km s$^{-1}$}\fi} 

\begin{document}

\title{JWST/MIRI detection of suprathermal OH rotational emissions: probing
the dissociation of the water by Lyman alpha photons near the protostar HOPS 370}

\author[0000-0001-8341-1646]{David A. Neufeld}
\affiliation{William H.\ Miller III Department of Physics and Astronomy, 
The Johns Hopkins University, 3400 North Charles Street, Baltimore, MD, US}

\author[0000-0002-3530-304X]{P. Manoj}
\affiliation{Department of Astronomy and Astrophysics, Tata Institute of Fundamental Research, Mumbai 400005, India}

\author[0000-0002-9497-8856]{Himanshu Tyagi}
\affiliation{Department of Astronomy and Astrophysics, Tata Institute of Fundamental Research, Mumbai 400005, India}

\author[0000-0002-0554-1151]{Mayank Narang}
\affiliation{Academia Sinica Institute of Astronomy \& Astrophysics, Taipei 10617, Taiwan}
\affiliation{Department of Astronomy and Astrophysics, Tata Institute of Fundamental Research, Mumbai 400005, India}

\author[0000-0001-8302-0530]{Dan M. Watson}
\affiliation{University of Rochester, Rochester, NY, US}

\author[0000-0001-7629-3573]{S. Thomas Megeath}
\affiliation{University of Toledo, Toledo, OH, US}

\author[0000-0001-7591-1907]{Ewine F. Van Dishoeck}
\affiliation{Leiden Observatory, Leiden University, P.O. Box 9513, 2300 RA Leiden, NL}
\affiliation{Max-Planck Institut f\"ur Extraterrestrische Physik, Garching bei München, DE}

\author[0000-0002-6447-899X]{Robert A. Gutermuth}
\affiliation{University of Massachusetts Amherst, Amherst, MA 01003, US}

\author[0000-0002-5812-9232]{Thomas Stanke}
\affiliation{Max-Planck Institut f\"ur Extraterrestrische Physik, Garching bei München, DE}

\author[0000-0001-8227-2816]{Yao-Lun Yang}
\affiliation{Star and Planet Formation Laboratory, RIKEN Cluster for Pioneering Research, Wako, Saitama 351-0198, Japan}

\author[0000-0001-8790-9484]{Adam E. Rubinstein}
\affiliation{University of Rochester, Rochester, NY, US}

\author[0000-0002-7506-5429]{Guillem Anglada}
\affiliation{Instituto de Astrof{\'i}sica de Andaluc{\'i}a, CSIC, Glorieta de la
Astronom{\'i}a s/n, E-18008 Granada, ES}

\author[0000-0002-1700-090X]{Henrik Beuther}
\affiliation{Max Planck Institute for Astronomy, Heidelberg, Baden Wuerttemberg, DE}


\author[0000-0001-8876-6614]{Alessio Caratti o Garatti}
\affiliation{INAF-Osservatorio Astronomico di Capodimonte, IT}

\author[0000-0001-5175-1777]{Neal J. Evans II}
\affiliation{Department of Astronomy, The University of Texas at Austin, Austin, TX 78712, US}

\author[0000-0002-6136-5578]{Samuel Federman}
\affiliation{University of Toledo, Toledo, OH, US}

\author[0000-0002-3747-2496]{William J. Fischer}
\affiliation{Space Telescope Science Institute, 3700 San Martin Drive, Baltimore, MD 21218, US}

\author[0000-0003-1665-5709]{Joel Green}
\affiliation{Space Telescope Science Institute, 3700 San Martin Drive, Baltimore, MD 21218, US}


\author[0000-0001-9443-0463]{Pamela Klaassen}
\affiliation{United Kingdom Astronomy Technology Centre, Edinburgh, GB}

\author[0000-0002-4540-6587]{Leslie W. Looney}
\affiliation{Department of Astronomy, University of Illinois, 1002 West Green St, Urbana, IL 61801, USA}
\affil{National Radio Astronomy Observatory, 520 Edgemont Rd., Charlottesville, VA 22903 US} 

\author[0000-0002-6737-5267]{Mayra Osorio}
\affiliation{Instituto de Astrof{\'i}sica de Andaluc{\'i}a, CSIC, Glorieta de la
Astronom{\'i}a s/n, E-18008 Granada, ES}

\author[0000-0002-4448-3871]{Pooneh Nazari}
\affiliation{European Southern Observatory, Karl-Schwarzschild-Strasse 2, 85748 Garching, Germany}


\author[0000-0002-6195-0152]{John J. Tobin}
\affil{National Radio Astronomy Observatory, 520 Edgemont Rd., Charlottesville, VA 22903 US}

\author[0000-0002-9470-2358]{Lukasz Tychoniec}
\affiliation{Leiden Observatory, Leiden University, P.O. Box 9513, 2300 RA Leiden, NL}

\author[0000-0002-0826-9261]{Scott Wolk}
\affiliation{Center for Astrophysics Harvard \& Smithsonian, Cambridge, MA, US}

\begin{abstract}

Using the MIRI/MRS spectrometer on {\it JWST}, we have detected pure rotational, 
suprathermal 
OH emissions from the vicinity of the intermediate-mass protostar HOPS 370 
(OMC2/FIR3).
These emissions are observed from shocked knots in a jet/outflow, and originate
in states of rotational quantum number as high as \rd{46} that 
possess excitation energies as large as $E_U/k = 4.65 \times 10^4$~K.
\rd{The relative strengths of the observed OH lines provide a \bx{  powerful
diagnostic} of the ultraviolet radiation field in a heavily-extinguished
region ($A_V \sim 10 - 20$) where direct UV observations are impossible.}
To high precision, the OH line strengths are consistent with a picture
in which the suprathermal OH states are populated following the photodissociation of
water in its $\tilde B - X$ band by ultraviolet radiation \rd{produced by fast 
($\sim 80\,\rm km\,s^{-1}$) shocks
along the jet}.  \rd{The observed dominance of emission from symmetric ($A^\prime$) OH
states over that from antisymmetric ($A^{\prime\prime}$) states provides a 
distinctive signature of this particular
population mechanism.  Moreover, the variation of intensity with rotational quantum number
suggests specifically that Ly$\alpha$ radiation is responsible for the photodissociation of
water, an alternative model with photodissociation by a 10$^4$~K blackbody being
disfavored at a high level of significance.}
Using measurements of
the Br$\alpha$ flux to estimate the Ly$\alpha$ production rate, we find that
$\sim \rd{4}\%$ of the Ly$\alpha$ photons are absorbed by water.  \rd{Combined
with direct measurements of water emissions in the $\nu_2 = 1 -0$ band, the OH 
observations promise to provide key constraints on future models for the diffusion of 
Ly$\alpha$ photons in the vicinity of a shock front.}

\end{abstract}


\section{Introduction}

In the realm of molecular astrophysics, one of the most remarkable results obtained
by {\it Spitzer} was the detection of highly suprathermal OH rotational emissions.
The observed transitions,
detected with the Short-Hi module of the Infrared Spectrometer (IRS)
toward the Herbig-Haro object HH 211 (Tappe et al.\ \rd{2008}), originate 
in pure rotational states with rotational quantum numbers $N$ as
high as 34 and energies as high as $E/k = 2.8 \times 10^4$~K.  They are naturally 
explained as the ``prompt emission'' that follows the photodissociation of water via
the $\tilde B - X$ band \bx{(also known as the ``second absorption band")} 
by radiation in the 114 -- 134~nm wavelength range;
this spectral region includes the strong Ly$\alpha$ line emitted by fast 
interstellar shocks.  This explanation is supported by both laboratory and theoretical
studies of water photodissociation through the $\tilde B - X$ band, which indicate that
OH states as high as $N=47$ can be populated (Harich et al.\ 2000, van Harrevelt \& 
van Hemert 2000).  \bx{Suprathermal OH emissions resulting from the photodissociation of 
water were subsequently observed in protostellar
disks with {\it Spitzer}: the protostellar disk of DG Tau, in particular, has been the subject of a
detailed analysis by Carr \& Najita (2014).}

{\it Spitzer} could not perform high-spectral resolution observations shortwards
of 10$\mu$m, the short wavelength cutoff of the Short-Hi module of the IRS, and at
shorter wavelengths the Short-Lo module on {\it Spitzer}/IRS provided a \rd{spectral 
resolving power, $\lambda/\Delta \lambda$}, of
only $\sim 60$, which was insufficient to detect suprathermal OH emissions.  By contrast, 
the  \rd{MIRI MRS spectrometer} on {\it JWST} provides coverage down to the OH band-head at 9.13~$\mu$m
(and below), yielding spectra with a \rd{spectral} resolving power $\sim 3000$.
This unique capability opens up the possibility of detecting suprathermal OH emission in the
9 -- 10 ~$\mu$m range, a possibility that has been realized in observations
of the Orion Bar reported very recently (Zannese et al. 2023), 
providing a powerful test of model predictions \bx{for the spectrum} of the OH prompt
emission (e.g. Tabone et al.\ 2021, hereafter T21).

In this {\it Letter}, we discuss {\it JWST}/MIRI observations of suprathermal OH emissions
in the vicinity of the protostar HOPS 370.  HOPS 370, a.k.a.\ OMC2/FIR3, is an
intermediate-mass \rd{Class 0/I 
protostar (Furlan et al. 2016)}.  
\rd{It is located north of the Orion Nebula in the OMC2 region of the 
integral shaped filament at an estimated distance of 392 pc 
(Kounkel et al. 2018, Tobin et al.\ 2020, hereafter T20).  Its central protostellar mass, 
determined from Keplerian motions, is $2.5\, M_\odot$, 
and its bolometric luminosity is $314\, L_\odot$ (T20).}
Extensive observations of HOPS 370
have been carried out with multiple observatories 
-- including {\it Herschel}, SOFIA, \bl{VLA}, ALMA, and now {\it JWST} -- and
together reveal an actively accreting protostar with a bipolar jet/outflow that is
orthogonal to a rotating disk of estimated mass 0.05 -- 0.1 $M_\odot$.  
It powers a large outflow traced in millimeter and far-IR lines, which suggests that 
it is in a state of rapid accretion (T20; Manoj et al.\ 2013; 
Gonz{\'a}lez-Garc{\'\i}a et al.\ 2016; 
Sato et al.\ 2023).  \bx{This outflow consists of both a wide-angle wind and 
a collimated jet, the latter containing shocks that are also seen in non-thermal 
radio emission (Osorio et al.\ 2017).} 
The orientation of the disk, with an estimated radius of 100 au, 
indicates that this source is observed at a high inclination angle of
$\sim 72^{\arcdeg}$ (T20; Federman et al.\ 2023, and references therein).
Luminous shocked knots in the northern outflow lobe are characterized by strong emissions
from a variety of molecules and atomic ions \rd{detected in our observations}, 
including H$_2$, H$_2$O, CO, OH, Fe$^+$, and Ne$^+$.  

In Section 2, we discuss the MIRI and NIRSpec observations carried out toward HOPS 370
and the methods used to reduce the data.  The resultant spectra and spectral line maps
are presented in Section 3, with particular emphasis on the suprathermal OH emissions
from the shocked knots.  The origin of those emissions is discussed in Section 4,
in the context of a model in which water is photodissociated by shock-produced
Ly$\alpha$ radiation.  \rd{A brief summary follows in Section 5.}

\section{Observations and data reduction}

The observations of HOPS 370 were performed as part of the
Cycle 1 medium GO program ``Investigating Protostellar Accretion (IPA),"
(PID 1802, Megeath et al. 2021), which carried out NIRSpec and MIRI 
IFU observations toward five protostars spanning \rd{five orders of
magnitude in luminosity} (see 
Federman et al.\ 2023).
A set of 2 x 2 mosaics was obtained with NIRSpec \rd{using the G395M/F290LP disperser-filter 
combination}, which provides
coverage of the 2.87 -- \rd{5.10} $\mu$m spectral region at 
a spectral resolving power $\lambda/\Delta \lambda \sim 1000$, and with 
all channels of the MIRI/MRS to provide complete mid-infrared 
coverage from 4.9 to 27.9 $\mu$m at spectral resolving power that
ranged from 1500 to 4000 (Jones et al.\ 2023).  The mosaicking was performed
with a 10$\%$ overlap and a 4-point dither pattern.  
The total \bl{observing} time was about 7.5 hours, including overheads. 
\rd{Further details of the
observing strategy have been presented by Narang et al. (2023) and 
Federman et al.\ (2023).}

\bl{For the reduction of NIRSpec IFU data, we utilized JWST pipeline version 1.9.5 and the 
JWST Calibration References Data System (CRDS) context version jwst$\_$1069.pmap. 
In our analysis, we identified hot pixels not captured by the JWST outlier detection step 
by applying a custom outlier detection algorithm specific to NIRSpec observations. 
More information on the NIRSpec data reduction and the custom flagging routine can be 
found in Federman et al. (2023).}

\bl{The MIRI MRS data reduction utilized JWST pipeline version 1.12.5 along with the JWST 
CRDS context version jwst$\_$1179.pmap. We used the standard Stage 1 JWST pipeline 
\textit{Detector1Pipeline} to reduce the MIRI MRS data starting from \textit{uncal} data.}

\bl{In the subsequent Stage 2 (\textit{Spec2Pipeline}), we performed pixel-by-pixel background 
subtraction using dedicated background observations. This process effectively 
removed all identified bad pixels, resulting in background-subtracted \textit{cal} products. 
However, we observed extended H$_2$ S(1) and H$_2$ S(2) emissions in the 
dedicated background observations, which led to reduced flux for these lines in the 
final data. Consequently, we repeated the \textit{Spec2Pipeline} without background subtraction. 
In this case, we encountered hot pixels in the detector data, which we removed using 
the \textit{VIP} package (Gomez Gonzalez et al., 2017; Christiaens et al., 2023).
Furthermore, we performed residual fringe correction during Stage 2 for both scenarios, 
with and without background subtraction.}

\bl{In Stage 3 (\textit{Spec3Pipeline}), the \textit{CubeBuildStep} was set to \textit{band} mode, 
generating separate FITS files for each channel and band. We also 
generated data cubes without dedicated background subtraction, with 
the outlier rejection function turned off in these cases.}

We measured and applied an astrometric offset calibration to the NIRSpec and MIRI IFU data 
to improve feature alignment and link the coordinates to the Gaia DR3 standard.  
The offset measurement process and listed offsets applied with uncertainties are 
presented in Federman et al.\ (2023).

Additional data reduction tasks were performed using
a suite of \texttt{Python} scripts we developed to (1) extract
spectra within a circular region of any specified position and radius;
(2) fit and subtract a continuum from the extracted spectra; 
(3a) fit Gaussian lines to continuum-subtracted spectra obtained
from task (2) above; or (3b) fit Gaussian lines with a first-order 
baseline at each IFU position and for each spectral line we targeted, thereby
enabling us to generate spectral line maps.

The second of these tasks \rd{(continuum fitting)}
was accomplished using a procedure that lacked any knowledge of the wavelengths
of expected spectral lines.  This ``zero-knowledge'' feature avoids the risk
of artificially creating spectral lines where lines are expected.
Here, for each spectral \bx{channel}, we fit a third-order 
polynomial to the fluxes measured within a 17-channel window centered on that spectral 
\bx{channel} 
(i.e. with 8 spectral channels on either side of the central 
one).  The fit was
optimized to achieve the best fit to any 10 of the 17 spectral channels in the window,
and the continuum flux value for the central channel was then assigned in accordance with
that fit.  For spectral regions where lines cover less than 7/17 $\sim 40\%$ of the
spectral samples, this procedure \rd{yields a reliable separation of 
the continuum} (including instrumental baseline ripples) from the lines.   For the third task, Gaussian fitting, we used the Levenberg–Marquardt 
algorithm; here, the line centroid and width were allowed to vary over a narrow range 
and the line intensity was allowed to vary freely, as were the continuum level and
slope for task (3b).

\section{Results}

The IFU data acquired toward HOPS 370 are extraordinarily rich, revealing 
literally hundreds of spectral lines with a signal-to-noise ratio adequate for mapping.
These data have and will be presented and discussed in series of papers, 
some already published (Federman et al. 2023; Rubinstein et al.\ 2023; \rd{Nazari et al.\ 2024}; Brunken et al.\ 2024) 
and some in preparation.  Here, we focus on the suprathermal OH lines and a small set of ancillary lines
that are directly relevant to their interpretation.

\subsection{Spectral line maps}

In Figure 1, we present maps of several spectral lines: 
a strong well-isolated water line within the $\nu_2 = 1 - 0 $ vibrational band;
the average of nine pure rotational lines of OH, with upper states with $N_U$ between
34 and 43\footnote{Here, we excluded the OH $N=37 - 36$ line, which lies very close 
to the \rd{much stronger} S(3) line of H$_2$}; the 
Br$\alpha$ line \rd{at 4.05~$\mu$m}, which traces the Ly$\alpha$ radiation responsible for the photodissociation
of water to produce suprathermal OH emissions; 
the \bl{$v=0-0$} S(3) line of H$_2$ \rd{at 9.66~$\mu$m}, one of eight pure 
rotational lines \bl{detected with MIRI/MRS} that may be used to
estimate the extinction toward the source; the [Fe II] 5.34 $\mu$m line,
a transition recently shown by Narang et al.\ (2023) to be an excellent tracer of collimated jets 
in another IPA target source, IRAS 16253-2429; 
and the [Ne II] 12.81 $\mu$m line, a signature of fast, ionizing shocks.
The maps are masked
in the vicinity of a bright continuum source in the south, \rd{MIPS 2301 (Megeath et al.\ 2012)},
where the line fits are 
unreliable.  The RA and Dec offsets are given
in arcsec relative to the ALMA source position (T20, green star): RA = 83.865142 deg,
Dec = --5.159561 deg (J2000).

Red circles near the lower right of each panel indicate the half power beam width
(HPBW) at the relevant wavelength, as determined
by the linear fit given by Law et al.\ (2023).
All these emissions peak roughly 0.8$^{\prime \prime}$ north of the ALMA source position, near 
the location of the shocked knots identified by Federman et al.\ (2023).  
The maps presented in Figure 1 exhibit a remarkable dynamic range: they are shown with 
a logarithmic stretch extending down to 0.1$\%$ of the peak intensity.  In units of 
$\rm 10^{-4}\, erg \, cm^{-2} \, s^{-1} sr^{-1}$, the peak 
velocity-integrated line intensities are 73, 1.14, 12.7, 36, 67, and 60 respectively for
the H$_2$O, OH, Br$\alpha$, H$_2$, [Fe II] and [Ne II] lines.  
For all the mapped lines other than the suprathermal OH emissions, a line is securely
detected in every spaxel right up to the edges of the mapped region.  
\bx{While the [Fe II] and [Ne II] fine structure emissions primarily trace a 
collimated bipolar jet, the H$_2$ emissions 
are much less strongly collimated (Federman et al.\ 2023)
and appear to trace a wide-angle wind. The Br$\alpha$, OH and
H$_2$O emissions show an intermediate degree of collimation.  
Velocity shifts, although smaller than the
instrumental linewidths, are clearly detected and
indicate that the northern jet is tilted towards us and the southern jet away from us.
They will be the subject of a future study.}

\subsection{OH suprathermal emission spectra}

In Figure 2, we present the 8.8 -- 13.4 $\mu$m spectra 
obtained toward the shocked knots
within the circular region indicated by the white circle in Figure 1. 
\rd{This aperture has a radius of 0.8$\arcsec$ and is centered at
a projected distance of 316 au from the protostar (ALMA position) on the OH emission
peak at offset $(\Delta \alpha {\rm cos}\delta, \Delta \delta) = (+0.1\arcsec,+0.8 \arcsec)$}.
The spectral region shown in Figure 2 covers \rd{24} securely-detected lines of OH, originating in states with
$N_U$ ranging from 23 to 46, along with 5 fine structure lines of [Ni II], [Co II], [Cl I]
and [Ne II], and two pure rotational lines of H$_2$, S(2) and S(3).

\includegraphics[angle=0,width=5.0 in]{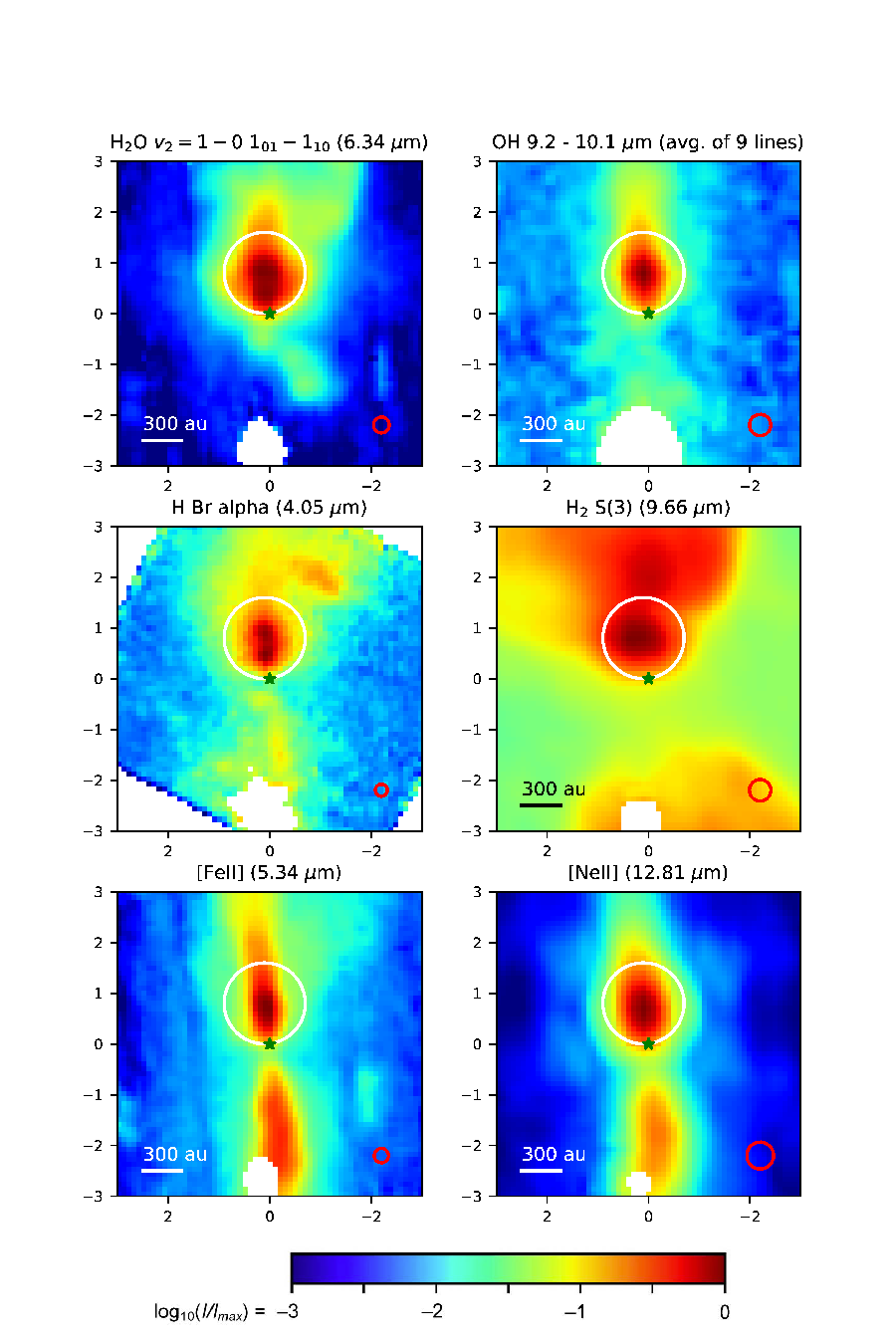}
\figcaption{Spectral line maps obtained toward HOPS 370, shown with a 
logarithmic stretch.  RA and Dec offsets are given
in arcsec relative to the ALMA source position (green star).  The white circle 
demarks a 0.8$^{\prime\prime}$ radius region centered on the shocked knot.  The maps are masked
near a bright continuum source in the south where the line fits are 
unreliable.  The red circles shown the beam size (HPBW).}

\includegraphics[angle=0,width= 5.8 in]{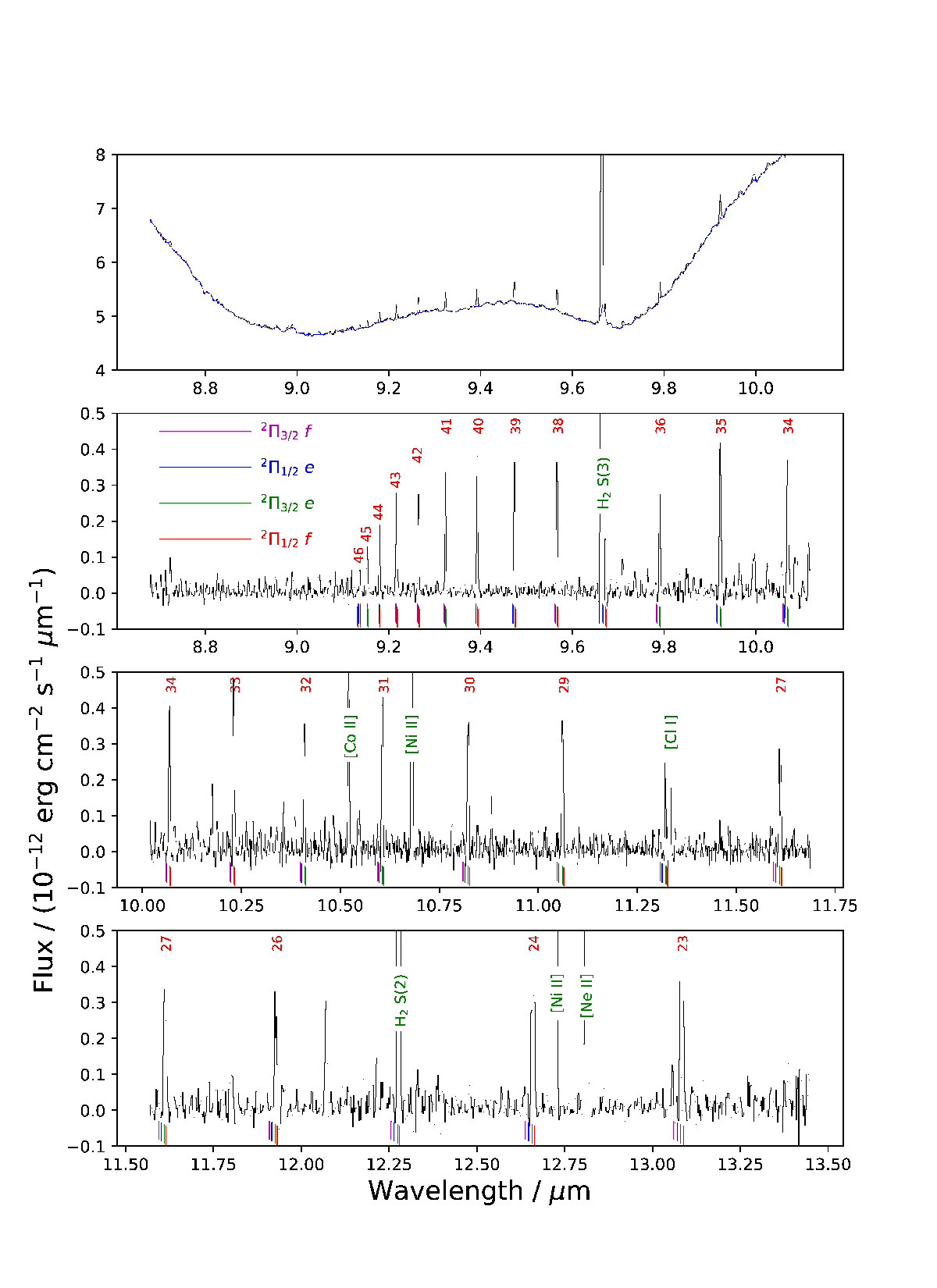}
\figcaption{8.8 -- 13.4 $\mu$m spectra obtained toward the shocked knots.  From top to
bottom: Band 2B spectrum with continuum fit in blue;
continuum-subtracted Band 2B, 2C, and 3A spectra.  Red numbers above the OH lines
indicate the value of $N_U$.}

From top to bottom, separate panels show the observed MIRI Band 2B (Channel 2 sub-band B) 
spectrum, with the
continuum fit (blue) obtained using the procedure for task (2) 
described in Section 2 above; and the continuum-subtracted Band 2B, 2C, and 3A spectra.
Colored \rd{vertical lines} at the bottom of the lower three panels show the positions of
suprathermal OH lines (Brooke et al.\ 2016), following the color coding 
indicated in the second panel from the top.  
\rd{Each rotational transition $N \rightarrow N-1$ is split into a quartet of
lines by the combined effects of lambda-doubling and spin-orbit coupling, and
the four lines are spectrally-resolvable by MIRI/MRS except at the highest values of $N$.}

Two of the four transitions connect so-called $A^\prime$ states, which are
symmetric with respect to reflection about the plane of rotation of the
molecule, and two connect antisymmetric $A^{\prime\prime}$ states.
(A further hyperfine splitting associated with the nuclear spin of H cannot
be resolved spectrally with MIRI/MRS for any of the observed transitions.)
The observed emission is completely dominated by intraladder transitions
involving symmetric $A^\prime$ states of OH (i.e.
the lower $e$ lambda doublets of the $^2 \Pi_{3/2}$ ladder and the lower $f$ 
lambda doublets of the $^2 \Pi_{1/2}$ ladder, shown with green and red lines.)

In Figure 3, we show zoomed spectra of the suprathermal OH lines (yellow histogram).  
Here, the black lines show Gaussian fits to each \rd{line}.  These were obtained with
the central wavelengths allowed to vary over a narrow range but with the wavelength 
separation of the two components fixed at the laboratory value and the flux ratio
of the two components fixed at unity.  The velocity scale is referenced
to the average wavelength of the two components.   
At the spectral resolution of MIRI, the 
separation of the $^2 \Pi_{3/2} (e)$ and $^2 \Pi_{1/2} (f)$ transitions is 
unresolved for the highest-$N_U$ lines
detected and fully-resolved for $N_U \le 23$.  \rd{Line positions are marked with vertical
lines for each component of the OH quartet, with the same color-coding as in
Figure 2: only the A$^\prime$ states (red and green) are detected.  Some transitions
with $N_U < 20$ are detected, but most lie in spectral regions where 
flux measurements are unreliable due to instrumental 
fringing.  They are not plotted here, and their
fluxes are not used in the analysis presented in Section 4.1 below.}

\includegraphics[angle=0,width= 7.8 in]{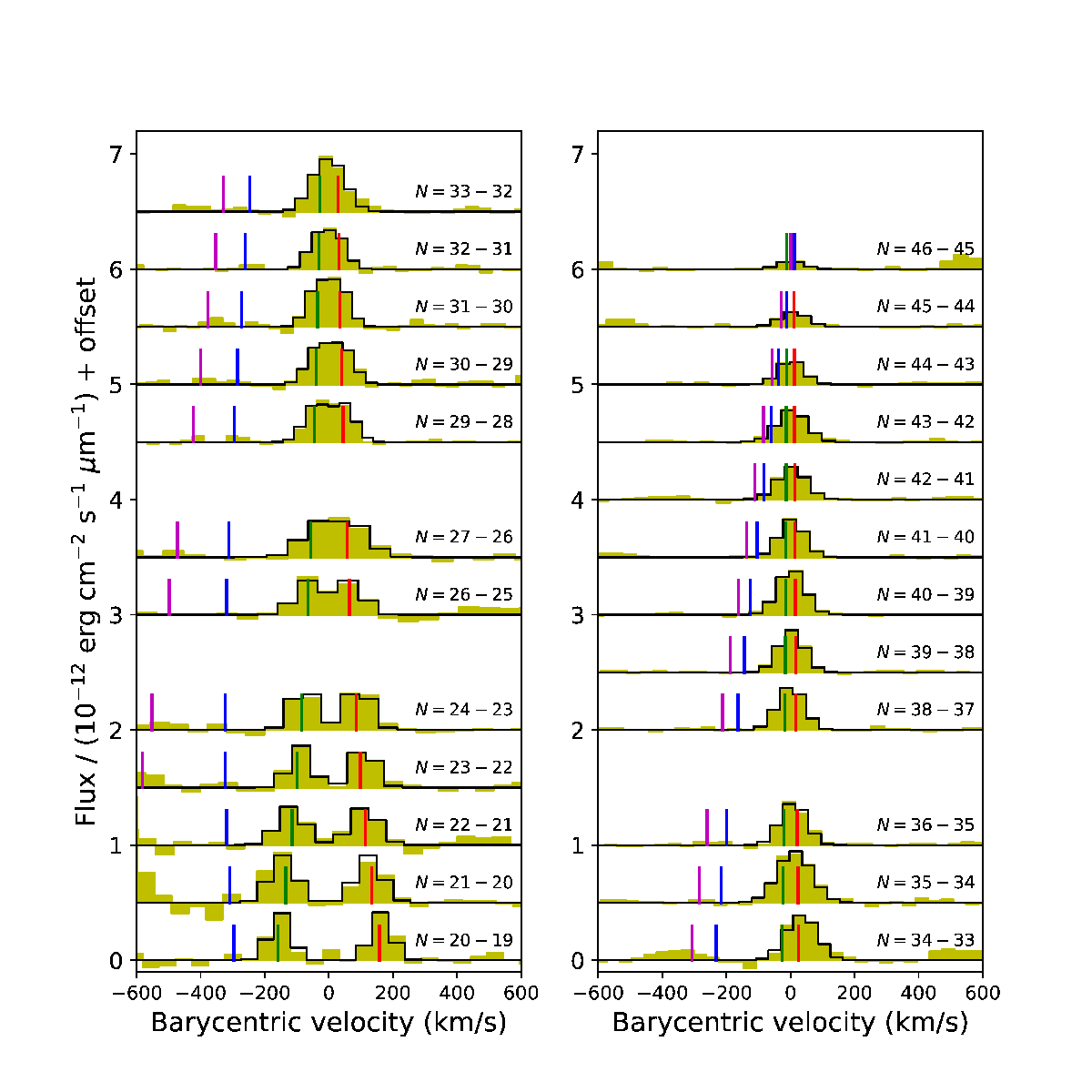}
\figcaption{Spectra of suprathermal OH lines observed toward the shocked knots.  
Yellow: observed spectrum.  Line positions are marked with vertical
lines for each component of the OH quartet, with the same color-coding as in
Figure 2.  Black histogram: Gaussian fit to the A$^\prime$ components
(see text).}

\subsection{$H_2$ rotational diagram and inferred extinction}

While the H$_2$ emissions from HOPS 370 will be discussed in detail 
in a future publication, their present relevance is simply in providing a valuable 
extinction estimate. Their usefulness for this purpose arises because the
S(3) line lies close to a local maximum in the extinction curve -- associated with 
the silicate absorption feature -- and therefore provides excellent leverage on 
the line-of-sight extinction.
Using the intensities of the S(1) through S(8) pure rotational lines of H$_2$, 
\rd{measured with MIRI/MRS}, we 
constructed the rotational diagram shown in Figure 4.  Here, we convolved the S(2) -- S(8) 
\rd{MIRI} maps with \rd{2D}-Gaussian kernels of the widths needed to degrade the 
\rd{spatial} resolution to a common
value for all lines.  We then obtained average \rd{intensities} for each line 
within the circular aperture indicated by the white circle in Figure 1.

Following Neufeld et al.\ (2006), for example, we fit the rotational diagram 
with the sum of two components \rd{each} in local thermodynamic equilibrium (LTE): a 
warm component at temperature $T_w$, with an aperture-averaged column density, $N_w$; 
and a hot component at temperature $T_h$, with an aperture-averaged column density, $N_h$.
These components were allowed to have separate ortho-to-para ratios, 
$\rm OPR_w$ and $\rm OPR_h$, 
yielding six free parameters to describe the rotational state of H$_2$.  The line-of-sight
extinction was treated as a seventh free parameter that was adjusted, along with the
other six, to optimize the fit (red and blue dashed curves).  {The best-fit values 
are indicated on Figure 4, and are typical of other protostellar outflows observed with {\it Spitzer}
(e.g.\ Neufeld et al.\ 2006).}
\footnote{\bx{The positive curvature of the rotational diagram, which we account for with a 
simple two-component model, suggests
that a range of gas temperatures is present (although the temperature distribution
need not be bimodal, e.g.\ Neufeld et al.\ 2009).  
The value of $\rm OPR_w$, lying significantly below 
the value of 3 expected in LTE at temperature $T_w$, is suggestive of 
transient heating in a shock wave; here, the OPR 
retains a fossil record of its cooler preshock state, there having been
insufficient time for it to reach equilibrium (Neufeld et al.\ 2006, and references
therein).}}

\includegraphics[angle=0,width= 6.5 in]{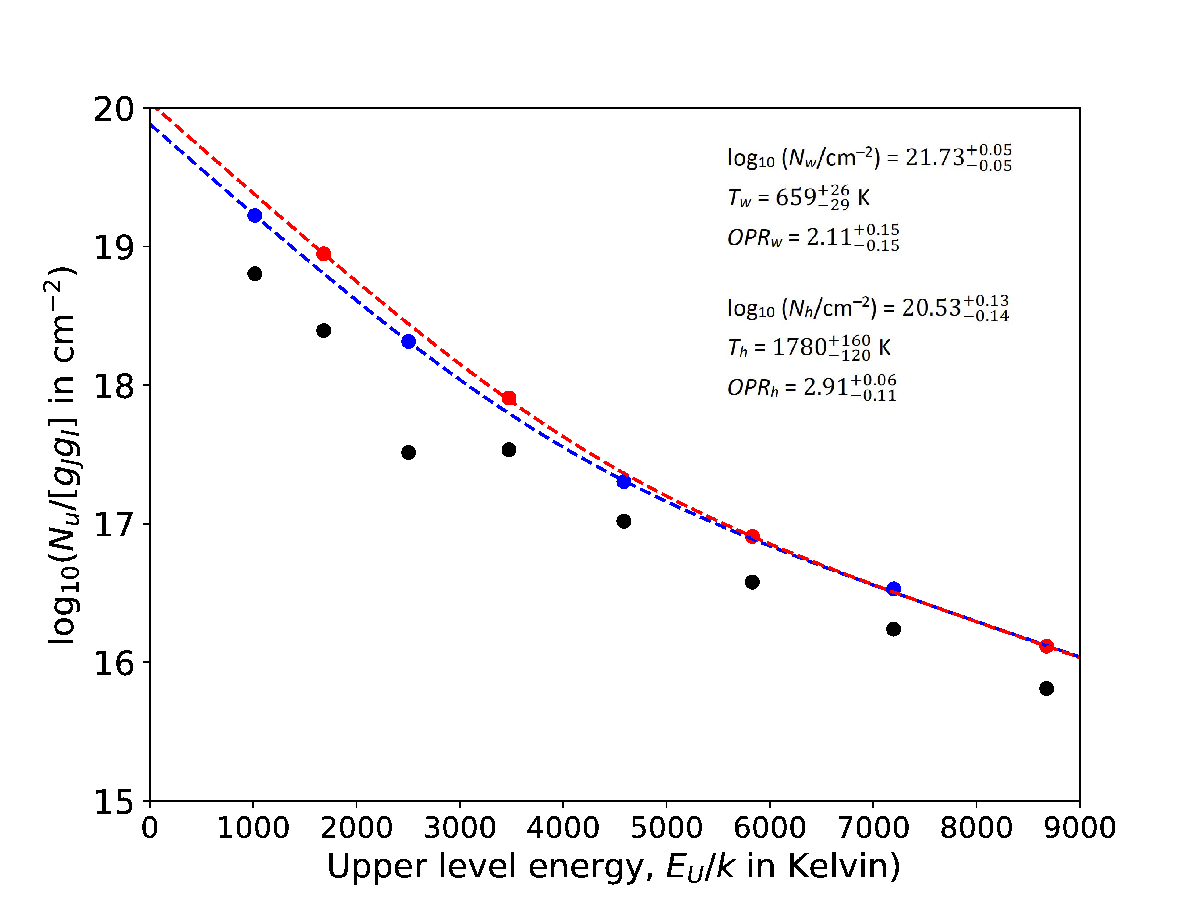}
\figcaption{H$_2$ rotational diagram obtained toward the shocked knots.  
Black points: no reddening correction.  
Blue and red points: reddening correction applied.  
Blue and red dashed lines: \rd{best fits to the rotational diagram 
for ortho and para-H$_2$.}}

Following Narang et al.\ (2024), we adopted the KPv5 extinction curve (Pontoppidan et al.\
2024) presented by Chapman et al.\ (2009), \rd{who cited a 2009 unpublished
study by K.\ Pontoppidan for its origin
and found that it provided the best fit to the mid-IR extinction 
and ice features observed in the {\it Spitzer} c2d program.} 
The black points indicate the column
densities in each rotational state inferred without any extinction correction, while
the red and blue points show the values inferred from the extinction-corrected line fluxes.
The best-fit extinction optical depth at 9.7$\mu$m is $\tau_{9.7}=\rd{1.84}$, and other fitting parameters
are specified in Figure 4.   

\rd{To evaluate the sensitivity of our conclusions to our choice
of extinction law and aperture size, we have also analysed the H$_2$ rotational emissions within
an aperture of radius 0.4$\arcsec$ instead of 0.8$\arcsec$ and for two additional 
mid-IR extinction laws that have appeared in the literature.  The results are discussed
in Appendix A, both as they pertain to the H$_2$ analysis discussed above and to the OH
analysis discussed below.  The relative OH lines fluxes 
favor the KPv5 extinction curve over the alternative extinction laws considered in 
Appendix A, but the primary conclusions of our study are similar regardless of which 
mid-IR extinction law or aperture size we adopt.}

\section{Discussion}

\subsection{Relative strengths of the suprathermal OH emission lines}

The high signal-to-noise ratio achieved in our observations of suprathermal OH emissions
facilitates a demanding test of theoretical models for their origin.  In Figure 5,
we plot the mean photon intensity, \rd{extinction-corrected with the KPv5 extinction curve}, 
as a function of $N_U$ 
(red crosses with bars showing the statistical errors).  Here, we excluded the OH
transitions with $N_U=37$, 28, and 25, which lie very close to the H$_2$ S(3), [Cl I] 11.33 $\mu$m, 
and H$_2$ S(2) lines, respectively. 

The results are in excellent
agreement with the predictions presented in Appendix D of T21,
which lists the number of OH line photons expected for each value of $N_U$
divided by the number of water photodissociations in the ${\tilde B} - X$ band. 
These calculations, which rest upon theoretical
calculations of the photodissociation dynamics (van Harrevelt \& van Hemert 2000)
and on experimental measurements (Harich et al.\ 2000),
were presented by T21 for four different radiation fields.  Those expected following 
water photodissociation by Ly$\alpha$ radiation are shown by the blue curve.  
There is only one free parameter in this comparison: an overall vertical scaling that is
proportional to the photodissociation rate within the beam.  If every available
UV photon led to a water photodissociation via the ${\tilde B} - X$ band,
the required UV photon intensity would be 
$\rd{I_{\rm UV} = \rm 1.67} \times 10^9 \rm  \, photons \, cm^{-2} \, s^{-1} sr^{-1}.$  
With 24 observed 
line intensities, the number of degrees of freedom here was $N_{\rm dof} = 23$.  

\bx{Our analysis here is closely-related to that of T21.  The only difference is
that we present the minimum possible photon intensity, $I_{\rm UV}$, that would account
for the absolute intensities of the observed OH emissions if every UV photon were absorbed 
locally by water.  This photon intensity is a factor 4$\pi$ smaller than the quantity 
$\Phi$ introduced by T21 and
referred to there as the column density of H$_2$O photodissociated per second.}

\includegraphics[angle=0,width= 7.0 in]{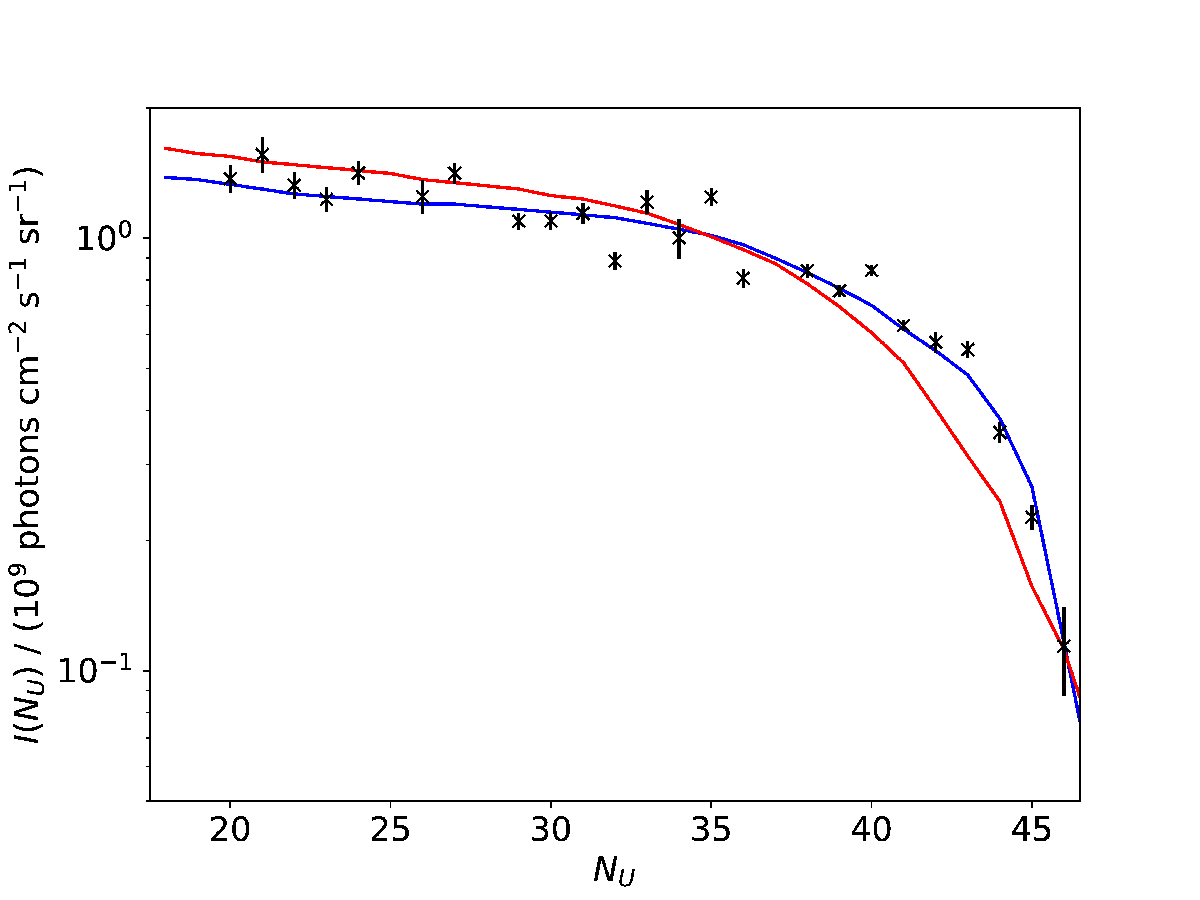}
\figcaption{OH photon intensity as a function of $N_U$.  Red points: observed intensities.
Also shown are the Tabone et al.\ (2021) predictions for H$_2$O photodissociation by
Ly$\alpha$ radiation (blue) and by a 10$^4$~K blackbody (red).}

The observational intensities clearly show systematic errors that are not
fully captured by the statistical error bars.  Assuming (1) that the predicted curve
for Ly$\alpha$ photodissociation (blue) represents the true behavior, (2) that the statistical and systematic 
errors both have Gaussian distributions with dispersions that
may be added in quadrature, and (3) that the fractional systematic
error has the same r.m.s., \bl{$\epsilon$}, for all lines,  
we adjusted \bl{$\epsilon$} to achieve a reduced $\chi^2$ of unity for the best-fit scaling.
The result was \bl{$\epsilon$ =} \rd{0.105}.

While the blue curve provides an excellent fit to the dependence of the line 
strengths on $N_U$, one aspect of the T21 predictions is in conflict 
with the observations.  Whereas T21 predict roughly equal populations in the 
\rd{symmetric ($^2 \Pi_{3/2} (e)$ and $^2 \Pi_{1/2} (f)$) and antisymmetric
states ($^2 \Pi_{1/2} (e)$ and $^2 \Pi_{3/2} (f)$)} of OH, the 
observations indicate that the symmetric \rd{$A^\prime$} states are
strongly favored; indeed, the antisymmetric \rd{$A^{\prime\prime}$} states are not
detected \bl{and are at least a factor $\sim 10$ less populated 
than the {$A^\prime$} states}.  \bx{Regardless of the relative rates at which
the symmetric and antisymmetric states are populated, the predictions presented 
in Appendix D of T21 are expected to apply to the total emission
in all four $N_U \rightarrow N_U -1 $ transitions (T21).}
 
This behavior, also noted in the recent paper of 
Zannese et al.\ (2023), 
is in fact entirely consistent with a recent theoretical study of the 
photodissociation process by Zhou et al.\ (2015), which indicates
a population ratio $A^\prime/A^{\prime\prime} \sim 40$ at the Ly$\alpha$ photon energy
(their Figure 7).  \rd{The astrophysical data thus provide a clear confirmation 
of the molecular physics.}  \bx{A less-pronounced difference ($\sim$ factor 2)
between the line fluxes for the $A^{\prime}$ and $A^{\prime\prime}$ transitions
had previously been measured by Carr \& Najita (2014)
in {\it Spitzer} observations of the 
protostellar disk in DG Tau.  These authors
discussed the effect in detail, with reference to two possible origins for 
OH: photodissociation of H$_2$O in the $\tilde B - X$ band, and chemical pumping following
formation via reaction of O($^1D$) with H$_2$.   The larger difference observed
in HOPS 370 may indicate that chemical pumping is relatively less important in this
source, at least for the $N_U \ge 20$ transitions discussed here.}

The $N_U$-dependence of the OH line intensities provides information about
the ultraviolet radiation field.  The red curve in Figure 5 shows
the predictions given by T21 (2021) 
for a blackbody radiation field at 10$^4$~K instead of
a Ly$\alpha$ radiation field.  These tend to overpredict the fluxes
for $N_U < 30$ relative to those for $N_U > 40$.  For \bl{$\epsilon$ =} \rd{0.105}, the 
minimum reduced $\chi^2$ for this case is $\chi^2_{\rm red} = \rd{2.98}$, 
implying that the blackbody radiation field 
is disfavored at the $[N_{dof}(\chi^2_{\rm red} - 1)]^{1/2} \sigma = \rd{6.7} \sigma$ significance
level.

As we did for the H$_2$ emissions discussed in Section 3.3, we have also analysed the OH 
emissions within an aperture of radius 0.4$\arcsec$ instead of 0.8$\arcsec$ and 
for two additional mid-IR extinction laws that have appeared in the literature.  
The results are discussed in Appendix A.

\subsection{Fraction, $f_w$, of Ly$\alpha$ photons absorbed by water}

The number of Ly$\alpha$ photons available to photodissociate water may be estimated
from the Br$\alpha$ flux observed within the circular aperture centered on 
the shocked knots.  \rd{In this analysis, we assume a geometry in which
the Br$\alpha$ and OH emissions are generated within the shocked knots and viewed directly rather 
than as a result of scattering, which is favored due their location in distinct shock knots.}
After degrading the resolution of the Br$\alpha$ map to
have the same HPBW as the OH lines, we obtain a value of
$\rm 1.4 \times 10^{-14} \, erg \, cm^{-2} \, s^{-1}$ for the Br$\alpha$ flux.  
If we apply an extinction correction 
assuming the value of $\tau_{9.7}$ obtained in 
Section 3.3 above, this corresponds to an intrinsic flux of 
$\rm 3.3 \times 10^{-14} \, erg \, cm^{-2} \, s^{-1}.$
\rd{As discussed in Appendix B, 
shock models appropriate for this source predict} typical Ly$\alpha$/Br$\alpha$ luminosity
ratios of $\sim \rd{900}$, a factor of 
several larger than the Case B recombination ratio 
because collisional excitation preferentially enhances Ly$\alpha$.

This would imply a Ly$\alpha$ flux within the aperture of 
$\rm \rd{3.0} \times 10^{-11} \, erg \, cm^{-2} \, s^{-1},$ 
or equivalently \rd{1.8} $\rm \, photons \, cm^{-2} \, s^{-1}$.
This corresponds to an aperture-averaged intensity of 
$\rm \rd{3.8} \times 10^{10}$ $\rm \, photons \, cm^{-2} \, s^{-1}\, sr^{-1}$,
a factor of $ \sim \rd{23}$ times as large as the minimum intensity of UV photons needed
to account for the OH line fluxes (Section 4.1 above).  Therefore, only a 
fraction $f_w = \rd{4.3}\%$ of the available
Ly$\alpha$ photons would need to be absorbed by water to explain the OH emission. 

\subsection{Interpretation of $f_w$}

Ly$\alpha$ photons are unlikely to \rd{travel far} without being
absorbed by dust or water.  
For Ly$\alpha$ radiation, we obtain a grain absorption cross-section per H nucleus
of $\sigma_{\rm abs}({\rm Ly} \alpha) = 1.9 \times 10^{-21}\, \rm cm^2$, 
adopting the wavelength dependence and albedo given by KPv5; 
here, the overall scaling 
was chosen to match the
average $N_H/A_J$ ratio of $5.6 \times \rm 10^{21} cm^{-2}\, mag^{-1}$  
determined by Vuong et al.\ 2003 from X-ray absorption observations in several nearby 
dense clouds\footnote{\rd{This scaling is also consistent with the standard $N_H/A_V$ 
ratio in diffuse molecular clouds (Bohlin et al.\ 1978), but yields a $N_H/A_K$ 
ratio $\sim 40\%$ smaller than the average values determined toward diskless
pre-main-sequence stars in Serpens and Orion by Winston et al.\ (2010)
and Pillitteri et al.\ (2013)}}.
The water photodissociation
cross-section for Ly$\alpha$ is $\rm 1.53 \times 10^{-17}\, \rm cm^2$ (Heays et al.\ 2017, 
and references therein), and thus
the ratio of the water absorption rate to the grain absorption rate for Ly$\alpha$ photons
is $R = 8 \times 10^3 \, x(\rm H_2O)$, where $x({\rm H_2O})= n({\rm H_2O})/n_{\rm H}$
is the water abundance relative to H nuclei.  The corresponding fraction
of Ly$\alpha$ photons absorbed by water is $f_w = R/(1 + R).$  
\rd{The estimate of $R$ given 
above is critically dependent on the (poorly known) properties of
grains in the outflow.  Indeed, it assumes that grains are present 
in protostellar outflows -- as suggested by Cacciapuoti et al.\ (2024) and 
references therein -- and moreover that their properties are similar to those in the 
dense interstellar medium.  The water abundance required to explain a given value
of $f_w$ scales linearly with the adopted value of $\sigma_{\rm abs}({\rm Ly} \alpha)$.} 

If $f_w = \rd{0.043}$ \rd{as determined in Section 4.2, and given the 
grain absorption cross-section assumed above}, 
required water abundance is $\rd{5} \times 10^{-6}$,
amounting to only $\sim 1\%$ of the gas-phase oxygen abundance  
\footnote{\rd{As noted in Appendix B, our estimate of the required water abundance
is quite strongly dependent upon the adopted grain properties:
an alternative and widely-used grain model in the literature yields a
value of only $\rd{1} \times 10^{-6}$}.}.
This is the average value, ${\bar x}(\rm H_2O)$,
encountered by the Ly$\alpha$ photons as they suffer repeated scatterings with
H atoms and execute a random walk prior to their eventual absorption.  In the region of
Ly$\alpha$ production, the gas is warm ($T \simgt 6000$~K) and/or ionized and the 
water abundance will be extremely small.  But if the photons escape the region where
they are produced without being absorbed by dust, then the water abundance could plausibly
exceed $10^{-4}$ if \rd{all oxygen nuclei were driven into gaseous water} and $R$ could exceed unity.  
In this scenario, the average water
abundance is less meaningful, and the quantity $f_w$ might primarily reflect the probability that 
a Ly$\alpha$ photon escapes the warm region where it is originally generated and enters
a region where water is abundant.  The transfer of Ly$\alpha$ radiation is a complex
process (e.g.\ Neufeld 1990) that can be profoundly affected by velocity shifts associated 
with shock waves (Neufeld \& McKee 1989).  We defer a detailed treatment of this process
to a future study.

\subsection{Lower limit on the water abundance from H$_2$O~$\nu_2= 1-0$ emissions}

The rovibrational water line \rd{map} shown in the upper left panel of Figure 1 shows just one
of several dozen emission lines detected in the H$_2$O~$\nu_2= 1-0$ band, which
collectively have a total equivalent width of $\sim \rd{0.120}$~$\mu$m.  \bx{Figure 6 shows
the 5.8 -- 7.0 spectral $\mu$m region that is dominated by these emission lines.}  
Unless the
density is extraordinarily high ($n_{\rm H} \simgt 10^9 \, \rm cm^{-3}$), these lines
are too strong to be produced by collisional excitation.
\bx{Colored symbols in Figure 6 show the line positions, with stars
denoting transitions of ortho-water and crosses denoting those of para-water. 
A color code (top left) indicates the minimum energies, $E_{\rm min}$, of the 
$v=0$ states that must be pumped radiatively to excite each transition.  We note 
here that $E_{\rm min}$ may be smaller than the energy, $E_L$, of the 
lower state of the observed rovibrational transition, since radiative pumping 
via a given transition may be followed by radiative decay in a different
transition of longer wavelength.
Roughly $90\%$ of the water emission emerges in transitions
that can be pumped radiatively out of the lowest 9 rotational states of water
(i.e. those with $J \le 2$ and $E/k < 160$~K).  This behavior suggests a low
rotational temperature within H$_2$O $v=0$ state, most likely because the 
states are subthermally populated, and supports the hypothesis of radiative
pumping.}
 
Although a full treatment of the
H$_2$O~$\nu_2= 1-0$ emissions is beyond the scope of the present study, 
we may
obtain a lower limit on the mean water abundance, $x({\rm H_2O})$, by assuming
that the observed lines are radiatively pumped by radiation from the protostar
and that the observed continuum is radiation from the protostar that has been
scattered by dust.  The equivalent width of the water lines is then
$$W_{\rm H_2O} = \sum {F_{\rm H_2O} \over F_c} \le   
\sum {\int \sigma_\lambda({\rm H_2O}) 
d\lambda \over \sigma_{\rm sca}} \biggl( {n_l({\rm H_2O}) \over n_{\rm H}} \biggr) \eqno(1)$$
where the sum is taken over all lines in the $\nu_2= 1-0$ band,
$W_{\rm H_2O}=\rd{0.120}\, \mu$m is the total equivalent width (summed over all ${\rm H_2O}$ lines), 
$F_{\rm H_2O}$ is the wavelength-integrated line flux for a given line,
$F_c$ is the continuum flux \rd{at the line wavelength}, $\sigma_{\rm sca}$ is the grain scattering cross-section per H 
nucleus, 
$n_l({\rm H_2O})$ is the number density of water molecules in the lower state,
and $\sigma_\lambda({\rm H_2O})$
is the ${\rm H_2O}$ cross-section for a given rovibrational line 
(per molecule in the lower state),
which has an integral over wavelength $\lambda$ given by $A_{ul}\lambda^4/(8 \pi c)$,
where $A_{ul}$ is the spontaneous radiative rate.  The equality in equation (1)
applies only if the pumping lines are optically-thin.

\includegraphics[angle=0,width= 7.2 in]{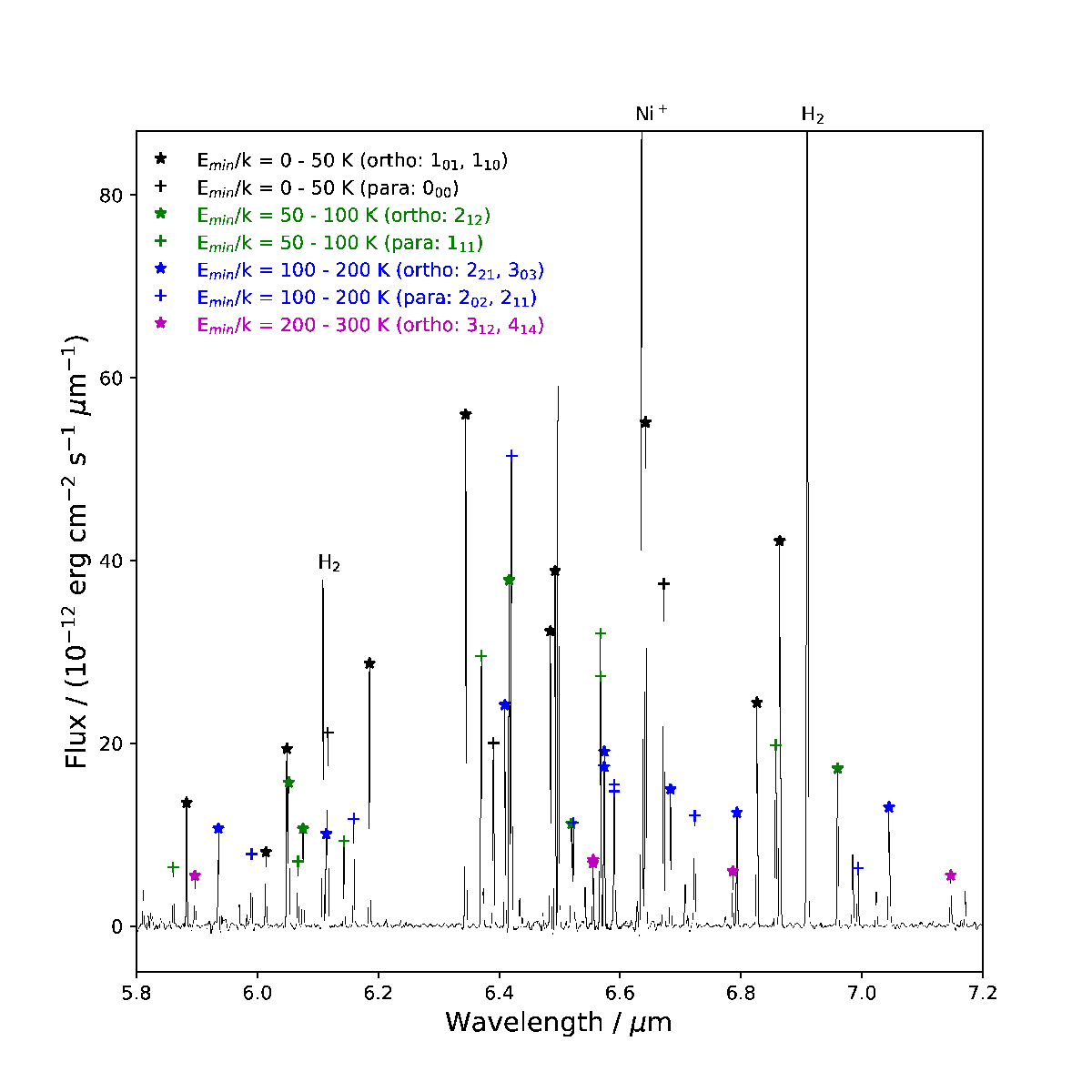}
\figcaption{\bx{Average 5.8 -- 7.0 $\mu$m spectrum obtained toward the shocked
knots, with rovibrational transitions of ortho- and para-water marked with stars and 
crosses.}} 

\rd{Because the KPv5 grain model suggests that $\sigma_{\rm sca}$ varies only slowly
over the band}, we may take $\sigma_{\rm sca}$ 
as a constant and approximate the sum of $A_{ul} \lambda^4 n_l({\rm H_2O})$ as 
$A_{band} \bar{\lambda}^4 n({\rm H_2O})$, where $A_{\rm band} = 24 \,\rm s^{-1}$ is
the total spontaneous radiation rate for the band and $\bar{\lambda} = 6.3\, \mu$m is the
average wavelength.  Given the grain scattering cross-section per H nucleus 
at $6.3\, \mu$m
implied by KPv5, $\sigma_{\rm sca}(6.3) = 1.2 \times 10^{-23}\, \rm cm^{-2}$,
we then obtain\footnote{A similar argument can be used to determine the CO abundance from observations
(Rubinstein et al.\ 2023) of the CO $v=1-0$ band: here, we obtain a total
equivalent width of 0.140 $\mu$m which implies a minimum CO abundance of 
$1.1 \times 10^{-4}$ relative to H nuclei.} 
$$x({\rm H_2O})={n({\rm H_2O}) \over n_{\rm H}} \ge {8 \pi c \sigma_s W_{\rm H_2O} 
\over A_{\rm band} \bar{\lambda}^4} = 3 \times 10^{-5}.\eqno(2)$$

Our water abundance of $3 \times 10^{-5}$ is a 
lower limit, and -- depending 
on the water linewidths -- would likely increase if optical depth effects are included.
But even this minimum estimate is a {\re factor 6} larger than the value needed 
to yield the inferred value of $f_w$ (Section 4.3 above).  This supports a picture in 
which most Ly$\alpha$ photons are absorbed by dust in a warm and/or ionized
zone very close to where they are created in a fast shock and only a minority
escape to the region of significant water abundance that is probed by the water
rovibrational emissions.
\rd{Like the estimate of ${\bar x}({\rm H_2O)}$ derived in section 4.3 above, this 
independent estimate of the minimum water abundance, derived
from $W_{\rm H_2O}$, is also dependent on the grain properties in the outflow; it 
scales linearly with the value adopted for $\sigma_{\rm sca}(6.3)$.  
If grains were depleted in the outflow (while maintaining the ratio
of $\sigma_{\rm sca}(6.3)$ to $\sigma_{\rm abs}({\rm Ly} \alpha)$),
both water abundance estimates would decrease proportionally.}
\bx{Future detailed analyses of the H$_2$O~$\nu_2= 1-0$ spectrum and how it varies
spatially will be needed to discriminate between the various mechanisms
that release water from ices into the gas-phase: these include thermal desorption, 
sputtering in shocks, and UV photodesorption.} 

\section{\rd{Summary}}

We have presented a study of OH in an outflow jet from the HOPS 370 protostar 
observed with MIRI and NIRSpec as part of the IPA program.

\noindent{1. We have detected pure rotational, 
suprathermal OH emissions from the vicinity of the intermediate-mass protostar HOPS 370 
(OMC2/FIR3).  These emissions are observed from shocked knots in a jet/outflow, and originate
in states of rotational quantum number as high as 46 that 
possess excitation energies as large as $E_U/k = 4.65 \times 10^4$~K.
Only symmetric $A^\prime$ states of OH are observed.}

\noindent{2. The relative OH line strengths are entirely consistent with a picture
in which the suprathermal OH states are populated following the photodissociation of
water in its $\tilde B - X$ band by Ly$\alpha$ radiation produced locally be a fast, ionizing 
shock.  Photodissociation by a blackbody radiation field at 10$^4$ K is found to provide
a significantly worse fit to the relative OH line strengths.}

\noindent{3. Using measurements of
the Br$\alpha$ flux to estimate the Ly$\alpha$ production rate
in shocked gas near HOPS 370, we find that
$\sim \rd{4}\%$ of the Ly$\alpha$ photons are absorbed by water.}

\noindent{4. The fraction of Ly$\alpha$ photons absorbed by water implies a mean
water abundance (relative to H nuclei) in the range 
${\bar x}(\rm H_2O) \sim 1 - 5 \times 10^{-6}$, the derived value depending upon the adopted
grain properties (Appendix B).   
This estimate is proportional to the grain absorption cross-section 
assumed
at the Ly$\alpha$ wavelength (121.6 nm), and
represents the average abundance within the  
region where Ly$\alpha$ photons scatter prior to being absorbed by dust or water.}

\noindent{5. Assuming that the H$_2$O $\nu_2$ band emissions observed from HOPS 370 are radiatively-pumped
and that the continuum is scattered light, we obtain a minimum water abundance in the range
$x_{\rm min} = 0.7 - 3 \times 10^{-5}$, the derived value depending upon the adopted
grain properties (Appendix A).  This minimum value is proportional to the grain 
scattering cross-section assumed at $6\,\mu$m, and would be exceeded if the pumping lines
are optically-thick.  It is a factor of several larger than ${\bar x}(\rm H_2O)$, suggesting that 
most Ly$\alpha$ photons are absorbed by dust in a warm and/or ionized
zone very close to where they are created in a fast shock and that only a minority
escape to the region of significant water abundance that emits the water
rovibrational emissions we observe.}

\noindent{6. Suprathermal OH emissions promise to help elucidate the processes whereby 
Lyman $\alpha$ radiation first escapes from fast shocks and then enters nearby 
water-rich surroundings
where water has been released from grain mantles by radiative heating or 
slower non-dissociative shocks, \bl{or produced in the gas-phase by neutral-neutral
reactions that are slow at low temperatures but rapid 
at the elevated temperatures attained behind shock fronts}.
Detailed models, beyond the scope of this {\it Letter}, will be needed to understand
the transfer of Ly$\alpha$ radiation and to fully model the rovibrational water 
emissions observed from HOPS 370.}

\acknowledgments{We thank the referee, Beno\^it Tabone, for
a very detailed and helpful review containing multiple suggestions
that improved this paper.
This work is based on observations made with the
NASA/ESA/CSA James Webb Space Telescope. The
data were obtained from the Mikulski Archive for Space
Telescopes at the Space Telescope Science Institute,
which is operated by the Association of Universities
for Research in Astronomy, Inc., under NASA contract
NAS 5-03127 for JWST. These observations are associated 
with program \#1802.  \bl{D.A.N.\ was supported by grant SOF08-0038 from USRA.
P.M.\ and H.T.\ acknowledge support of the Department 
of Atomic Energy, Government  of India, under Project Identification 
No.\ RTI 4002.}  Support for S.F., A.E.R.,
S.T.M., R.G., W.F., J.G., J.J.T. and D.W. in program \#1802
was provided by NASA through a grant from the Space
Telescope Science Institute, which is operated by the
Association of Universities for Research in Astronomy,
Inc., under NASA contract NAS 5-03127. A.C.G. has
been supported by PRIN-MUR 2022 20228JPA3A “The
path to star and planet formation in the JWST era
(PATH)” and by INAF-GoG 2022 “NIR-dark Accretion
Outbursts in Massive Young stellar objects (NAOMY)”.
G.A. and M.O., acknowledge financial support from
grants PID2020-114461GB-I00 and CEX2021-001131-
S, funded by MCIN/AEI/10.13039/501100011033. Y.-
L.Y. acknowledges support from Grant-in-Aid from the
Ministry of Education, Culture, Sports, Science, and
Technology of Japan (20H05845, 20H05844, 22K20389),
and a pioneering project in RIKEN (Evolution of Matter in the Universe). 
W.R.M.R. is grateful for support from
the European Research Council (ERC) under the European Union’s 
Horizon 2020 research and innovation programme (grant agreement 
No. 101019751 MOLDISK).
\bx{All the data presented in this article were obtained from the Mikulski Archive 
for Space Telescopes (MAST) at the Space Telescope Science Institute. 
The specific observations analyzed can be accessed 
via \dataset[10.17909/3kky-t040]{https://doi.org/10.17909/3kky-t040}} 
}

\vfill\eject
\appendix

\section{Dependence on adopted extinction curve and aperture size}

We have evaluated the sensitivity of our conclusions to our choice
of extinction law and aperture size.  Results are presented in Table 1 for
six cases.  We considered three different extinction curves -- denoted KP
(KPv5), WD (Weingartner and Draine, 2001, with the modifications described in
Draine 2003); and MM (McClure 2009) -- and two different aperture 
radii (0.8$\arcsec$ and 0.4$\arcsec$, denoted by 1 and 2).  The standard model, adopted in
the main text, is KP1 (KPv5 extinction curve with the 0.8$\arcsec$ radius aperture).

Different rows in Table 1 show the values obtained for key parameters for all six cases.
The first row below the horizontal line 
lists the  optical depth at 9.7$\,\mu$m, $\tau_{9.7}$,
derived from our fit to the H$_2$ lines.
It ranges from 1.53 to 3.12, with the MM extinction law (which is significantly less dominated 
by the silicate peak) requiring the largest $\tau_{9.7}$ and the WD extinction law requiring the
smallest, but shows very little variation with the adopted aperture size. 
The second row lists, $I_{UV}$, the required UV photon intensity
if every available UV photon led to a water photodissociation via 
the ${\tilde B} - X$ band.  In determining $I_{UV}$, the OH lines were extinction-corrected
using the specified extinction curve and the corresponding value of $\tau_{9.7}$, and
Ly$\alpha$ was assumed to be responsible for water photodissociation.  
As discussed in Section 4.1, we assumed equal fractional systematic errors, \bl{$\epsilon$}, 
for each OH flux measurement and adjusted \bl{$\epsilon$} to yield a reduced $\chi^2$ of unity
when comparing case KP1 and KP2 fluxes with the predictions for Ly$\alpha$ photodissociation.
The third row lists the reduced $\chi^2$ obtained for each of the six cases.  
The fourth row lists the corresponding values obtained for a 10$^4$~K 
blackbody radiation field instead of Ly$\alpha$.
The fifth and sixth rows indicate the significance with which each case is disfavored
relative to KP1 or KP2 with photodissociation by Ly$\alpha$.  The values plotted here
indicate (1) that the KP extinction law yields a significantly better fit to the data
than either WD or MM; (2) for any extinction curve, assuming photodissociation by 
Ly$\alpha$ radiation yields a significantly better fit to the data than does 
assuming photodissociation by a $10^4$~blackbody.  

The seventh row in Table 1 lists the extinction-corrected (ec) Br$\alpha$ intensity, 
$I(Br\alpha)_{\rm ec}$ for each of the six cases.  The next row lists the
corresponding values of $f_w$, the fraction of Ly$\alpha$ photons absorbed by dust.
The quantity $f_w$ was computed for an assumed Ly$\alpha$/Br$\alpha$ ratio of 900 
(see Appendix B) and is proportional to $I_{UV}/I(Br\alpha)_{\rm ec}$.  Although
$I_{UV}$ varies over a factor of 4 -- 5 as different extinction curves are adopted,
$f_w$ varies by less than a factor 2.  This is because there is some degree of
cancellation between the effects on $I_{UV}$ and $I(Br\alpha)_{\rm ec}$.  The MM extinction
curve, for example, requires a significantly larger $\tau_{9.7}$ leading to a 
significantly larger $I_{UV}$: but because the opacity is less strongly peaked at silicate
feature, $I(Br\alpha)_{\rm ec}$ also increases by a factor 2.   

The conclusions of our sensitivity analysis are (1) similar results are obtained for
both aperture sizes; (2) as the origin of the suprathermal OH emissions, 
water photodissociation by Ly$\alpha$ radiation is robustly favored
over photodissociation by a $10^4$~K blackbody; (3) the KPv5 extinction is favored
over the other two extinction laws considered here; and (4) the parameter $f_w$ lies in the
range 2.6 - 4.3 $\%$ for any of the extinction laws we considered.

Rows 9 - 13 are relevant to the water abundances discussed in Sections 4.3 and 4.4.
Here, not only is the relative extinction at different wavelengths of relevance, but
so too is the grain albedo and the ratio of extinction to column density, $N_{\rm H}$.
Both these parameters are available for the KP and WD extinction laws but not for MM.
Row 9 lists the grain absorption cross-section per H nucleus, $\sigma_{\rm abs}({\rm Ly} \alpha)$,
given by each extinction law.  The average water abundance, ${\bar x}(\rm H_2O)$, 
needed to account for $f_w$ is listed in Row 10 (see Section 4.3).   Row 11 lists the
grain scattering cross-section, $\sigma_{\rm sca}(6.3)$, needed for the 
analysis presented in Section 4.4, while Row 12 lists the total $\nu_2 = 1-0$
equivalent width, $W_{\rm H2O}$, for each aperture.  The resultant minimum water abundances, 
$x_{\rm min}(\rm H_2O)$ (see Section 4.3), are given in Row 13.  

The results presented here for ${\bar x}(\rm H_2O)$ and $x_{\rm min}(\rm H_2O)$ are
clearly more sensitive to uncertainties in the dust properties. Nevertheless, the conclusion
that ${\bar x}(\rm H_2O) < x_{\rm min}(\rm H_2O)$ appears to be robust, with the
implications for Ly$\alpha$ radiative transfer discussed in Section 4.4.

\begin{deluxetable}{llcccccc}
\tabletypesize{\footnotesize}
\tablecaption{Results for different extinction laws and aperture sizes}

\tablehead{Row & Quantity & KP1 & WD1 & MM1 & KP2 & WD2 & MM2  \\
						&&  0.8$\arcsec$ & 0.8$\arcsec$ & 0.8$\arcsec$ & 0.4$\arcsec$ & 0.4$\arcsec$ & 0.4$\arcsec$}
\startdata
1 & $\tau_{9.7}$ 											& 1.84 & 1.53 & 3.12 & 1.85 & 1.53 & 3.05 \\
2 & $I_{UV}$ (10$^9$ photons cm$^{-2}$ s$^{-1}$ sr$^{-1}$)	& 1.67 & 1.10 & 5.11 & 3.75 & 2.42 & 1.07 \\
3 & Reduced $\chi^2$ (w.r.t.\ Ly$\alpha$ model) 			& 1.00 & 1.63 & 1.55 & 1.00	& 2.65 & 2.33 \\
4 & Reduced $\chi^2$ (w.r.t.\ 10$^4$ K BB model) 			& 2.98 & 6.13 & 4.60 & 3.74 & 8.40 & 7.30 \\
5 & Significance$^a$ (for Ly$\alpha$ model) 		 		& 0.0$\sigma$ & 4.4$\sigma$ & 3.6$\sigma$  & 0.0$\sigma$  & 6.2$\sigma$  & 5.5$\sigma$ \\
6 & Significance$^a$ (for 10$^4$ K BB model) 				& 6.7$\sigma$ & 10.9$\sigma$ & 9.1$\sigma$  & 7.9$\sigma$  & 13.0$\sigma$ & 12.0$\sigma$ \\
7 & $I(Br\alpha)_{\rm ec}$ (10$^{-3}$ erg cm$^{-2}$ s$^{-1}$ sr$^{-1}$)	& 0.71 & 0.72 & 3.64 & 1.49 & 1.52 & 7.22 \\
8 & $f_w$ 													& 0.043 & 0.028 & 0.025 & 0.046 & 0.029 & 0.027 \\
9 & $\sigma_{\rm abs}({\rm Ly}\alpha)$ ($10^{-21}\,\rm cm^2$) & 1.73 & 0.70 & N/A & 1.73 & 0.70 & N/A \\
10 & ${\bar x}(\rm H_2O)$ 									& $5.1 \times 10^{-6}$ & $1.3 \times 10^{-6}$ & N/A &$5.1 \times 10^{-6}$ & $1.3 \times 10^{-6}$ & N/A \\
11 & $\sigma_{\rm sca}(6.3)$ ($10^{-23}\,\rm cm^2$)			& 1.26 & 0.29 & N/A & 1.26 & 0.29 & N/A \\ 
12 & $W_{\rm H2O}$ ($\mu m$) 								& 0.120 & 0.120 & 0.120 & 0.168 & 0.168 & 0.168 \\
13 & $x_{\rm min}(\rm H_2O)$								& $3.0 \times 10^{-5}$ & $7.0 \times 10^{-6}$ & N/A & $4.2 \times 10^{-5}$ & $1.0 \times 10^{-5}$ & N/A \\
\enddata
\tablenotetext{a}{Significance with which a given model and extinction curve is disfavored
relative to the model with Ly$\alpha$ photodissociation and the KP extinction curve.}
\end{deluxetable}

\section{Shock model predictions for L\lowercase{y}$\alpha$/B\lowercase{r}$\alpha$}

We have used publicly-available the MAPPINGS V shock model (Sutherland \& Dopita 2017; Sutherland et al.\ 2018) 
to estimate 
the Ly$\alpha$/Br$\alpha$ luminosity ratio within the shocked region where suprathermal OH emissions
were detected.  \rd{The upper states of these lines may be populated 
both following recombination of H$^+$ and by direct colllisional excitation of 
neutral hydrogen from the ground state.}
The Ly$\alpha$/Br$\alpha$ ratio can significantly exceed the Case B recombination
value $\sim 300$, particularly for lower velocity shocks where collisional excitation
is most important, so the use of shock model predictions is important here.

We ran a grid of models with preshock densities, $n_0$, spanning the 
range $10^{-1}$ to $10^{5.5}$ H nuclei per cm$^{-3}$ in steps of 0.1 dex; and with shock 
velocities, $v_s$, spanning the range 30 to 220 km~s$^{-1}$ in steps of 5 km~s$^{-1}$.
The preshock ionization state was determined self-consistently.  The
preshock magnetic field was taken as $0.5\,(n_0/\rm cm^{-3})^{1/2}\, \mu G$, and undepleted
solar abundances were adopted.  The collision strengths for [Fe II] fine structure 
transitions were updated to the values given in the recent study of Tayal \& Zatsarinny
(2018).

Figure 7 shows contours of the predicted Ly$\alpha$/Br$\alpha$ luminosity ratio
in the plane of $v_s$ and log$_{10}n_0$.  Here, the red, cyan and blue contours show
where the Ly$\alpha$/Br$\alpha$ ratio is predicted to be 1000, 1500, and 2000, and black 
contours show intermediate values spaced by 100.  The [Ne III]~15.6~$\mu$m to 
[Ne II]~12.8 $\mu$m flux ratio is an excellent tracer of shock velocity.  The observed,
extinction-corrected
value of 0.026 is obtained for shock parameters lying along the locus marked with the 
green solid curve.  The green band indicates the region where the predicted value lies
within a factor 1.5 of that observed, with the dotted/dashed boundaries applying to 
larger/smaller line ratios.  As a probe of the preshock density, we have considered
the [Fe II]~17.9~$\mu$m to [Fe II]~5.3~$\mu$m flux ratio, which has an observed 
extinction-corrected value of 4.17.  The magenta curves and magenta band represent analogous results
for the [Fe II] line ratio.  The constraint on density is less tight, as indicated \bl{by} the 
width of the magenta band, and must be considered less reliable because recent
independent estimates of the collision strengths show significant differences.
Nevertheless, the intersection of the green and magenta solid lines suggest
that Ly$\alpha$/Br$\alpha$ ratio of 900 is consistent with these diagnostic line
ratios. For any preshock density in the range 10$^2$ to 10$^5$~cm$^{-3}$, the flux
[Ne III]~15.6~$\mu$m to [Ne II]~12.8 $\mu$m flux ratio {\it alone} suggests a 
Ly$\alpha$/Br$\alpha$ ratio in range 700 -- 1050. 

\includegraphics[angle=0,width= 6.5 in]{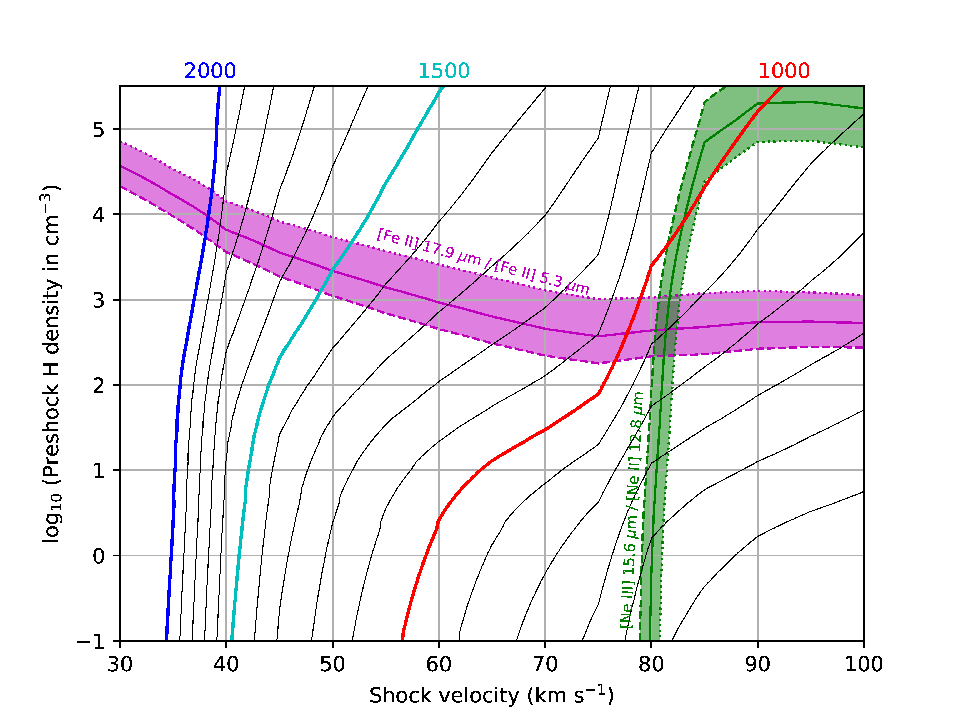}
\figcaption{Contours of the predicted Ly$\alpha$/Br$\alpha$ luminosity ratio
in the plane of $v_s$ and log$_{10}n_0$. Red, cyan and blue contours:
Ly$\alpha$/Br$\alpha$ = 1000, 1500, and 2000.  Green band: region where the predicted 
[Ne III]~15.6~$\mu$m to [Ne II]~12.8 $\mu$m flux ratio lies
within a factor 1.5 of that observed.  Magenta band: region where the predicted 
[Fe II]~17.94~$\mu$m to [Fe II]~5.34 $\mu$m flux ratio lies within a factor 1.5 of that observed.}

\vfill\eject

\end{document}